\documentclass[sigconf]{acmart}

\AtBeginDocument{%
  \providecommand\BibTeX{{%
    \normalfont B\kern-0.5em{\scshape i\kern-0.25em b}\kern-0.8em\TeX}}}

\setcopyright{none}
\copyrightyear{.}
\acmYear{.}
\acmDOI{.}

\acmConference{.}
\acmBooktitle{.}
\acmPrice{.}
\acmISBN{.}
\renewcommand\footnotetextcopyrightpermission[1]{} 
\usepackage{url}
\usepackage{tabularx}
\usepackage{xcolor}
\usepackage[binary-units]{siunitx}
\usepackage{setspace}
\usepackage{acronym}
\usepackage{rotating}
\usepackage{subcaption}
\usepackage{units}
\usepackage{siunitx}
\usepackage{multirow}
\usepackage{listings}
\usepackage{pgfplots}
\usepackage{pgf-pie}
\newacro{VM}{virtual machine}
\newacro{MPU}{memory protection unit}

\begin{document}

\title{Femto-Containers: DevOps on Microcontrollers with Lightweight Virtualization \& Isolation for IoT Software Modules}

\author{Koen Zandberg}
\affiliation{%
  \institution{Inria, France}
 \country{.}
}

\author{Emmanuel Baccelli}
\affiliation{%
  \institution{Inria, France}
  \country{Freie Universit\"at Berlin, Germany}
}

\renewcommand{\shortauthors}{.}

\begin{abstract}
Development, deployment and maintenance of networked software has been revolutionized by DevOps,
which have become essential to boost system software quality and to enable agile evolution.
Meanwhile the Internet of Things (IoT) connects more and more devices which are not covered by DevOps tools:
low-power, microcontroller-based devices.
In this paper, we contribute to bridge this gap by designing Femto-Containers,
a new architecture which enables containerization, virtualization
and secure deployment of software modules embedded on microcontrollers over low-power networks.
As proof-of-concept, we implemented and evaluated Femto-Containers on popular microcontroller architectures (Arm Cortex-M, ESP32 and RISC-V), using eBPF virtualization, and RIOT, a common operating system in this space.
We show that Femto-Containers can virtualize and isolate multiple software modules, executed concurrently, with very small memory footprint overhead (below 10\%) and very small startup time (tens of microseconds) compared to native code execution.
We show that Femto-Containers can satisfy the constraints of both
low-level debug logic inserted in a hot code path, 
and high-level business logic coded in a variety of common programming languages.
Compared to prior work, Femto-Containers thus offer an attractive trade-off in terms of memory footprint, energy consumption, agility and security.

\end{abstract}

\maketitle

\section{Introduction}

An estimated 250 billion microcontrollers are in use today~\cite{250billionMCU}.
An increasing percentage of these microcontrollers are networked and take part in distributed cyber-physical systems and the Internet of Things (IoT) we increasingly depend upon.

With the availability of low-power operating systems~\cite{hahm2015operating} and network stacks (e.g. 6LoWPAN), low-power IoT software has made giant leaps forward ; but fundamental gaps remain compared to current practices for networked software.
In fact, current state-of-the-art for managing, programming, and maintaining fleets of low-power IoT devices resembles more PC system software workflow from the 90s than today's common software practices: simplistic application programming interfaces (APIs) offer basic performance and connectivity, but no additional comfort.

However, since the 90s, networked software was revolutionized many times over. Networked software has entered the age of agility. DevOps~\citep{bass2015devops} drastically shortened software development/deployment life cycles to provide continuous delivery of higher software quality. Additional layers providing cybersecurity, flexibility and scalability thus became crucial : ubiquitous script programming (e.g., Python, Javascript), light-weight software containerization (e.g., Docker), hypervisors and software virtualization, deployment and management tools for swarms of virtualized software instances (e.g., Kubernetes or AWS), and frameworks for decentralizing system software updates, development and maintenance on platforms such as Linux, Android, iOS, Windows etc.

In such a context, low-power IoT devices are the new ‘weakest link’ within distributed cyber-physical systems upon which relies IoT and related services.
Indeed, state-of-the-art DevOps mechanisms are not commonly applicable on low-power devices : they are either not applicable on microcontrollers (e.g., Docker), too prohibitive in terms of hosting engine memory resource requirements (e.g., OS virtualization or standard Java virtual machines), or restricted to very specific use cases (e.g., JavaCard).
This lackluster create bottlenecks which severly impact both flexibility and cybersecurity in IoT.

A key question emerges: can we provide new concepts adequate for software containerization and rapid deployment on swarms of IoT devices, combining agility, low-power consumption and cybersecurity?
The goal we pursue in this paper is to explore practical solutions for software virtualization, containerization and deployment applicable to fleets of low-power, connected, microcontroller-based IoT devices.

{\bf Contributions --}
In this paper, the work we present mainly consists in the following:
\begin{itemize}
\item we survey existing techniques for process isolation \& virtualization for microcontrollers;
\item we design Femto-Containers, a novel secure DevOps architecture for constrained IoT devices;
\item we implement Femto-Containers based on eBPF virtualization, and small containers hosted in a common low-power IoT operating system (RIOT);
\item we evaluate the performance of Femto-Containers in a variety of use cases, on popular microcontroller architectures (Arm Cortex-M, ESP32 and RISC-V);
\item we compare Femto-Containers with prior work based on WebAssembly, JavaScript (RIOTjs) and Python (microPython). We show Femto-Containers offer an attractive trade-off in terms of memory footprint, energy consumption, and security.
\end{itemize}

\section{Scenarios \& Threat Model}
\label{sec:case-study}
\label{sec:isolationmodel}
\begin{figure}[t]
  \centering
\includegraphics[width=\columnwidth]{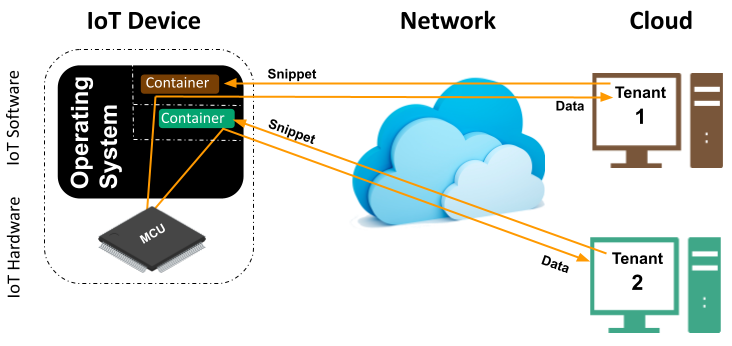}
  \caption{DevOps use-cases on IoT microcontrollers.}
  \label{fig:femto-containers-use-cases}
\end{figure}

As low-power embedded IoT software complexifies on various devices, it becomes necessary for security (and sometimes also privacy) reasons to delegate maintenance and updates of different parts of the embedded software to distinct entities with limited mutual trust (as described in~\cite{thomas2018design} for instance).
Furthermore, enabling on-the-fly and safe remote instrumentation of already-deployed software is essential to maintain cyberphysical systems involving low-power IoT devices.
We thus consider different categories of use-cases, depicted in \autoref{fig:femto-containers-use-cases}:
\begin{enumerate}
  \item Use-case 1: Hosting and isolating some high-level business logic, updatable on-demand remotely over the low-power network. The execution of this type of logic  is typically periodic in nature, and has loose (non-real-time) timing requirements.
  \item Use-case 2: Hosting and mutually isolating several virtual machines, managed by several different tenants.
  \item Use-case 3: Hosting and isolating some debug/monitoring code applications at low-level, inserted and removed on-demand, remotely, over the network. Comparatively, this type of logic is short-lived and exhibits stricter timing requirements.
\end{enumerate}

{\bf Threat Model.} We consider both malicious tenants which can deploy malicious containers and malicious clients which can maliciously interact with deployed containers.

\emph{Malicious Tenant}:
While a tenant has to work within the permissions granted by the host system, it can make free use of the granted resources.
A tenant could provide a vulnerable or malicious application code for execution inside a container.
To protect against this, the host system must ensure that hosted applications are constrained in the resources they can access and have a fair share of the processing time and network bandwidth. 

\emph{Malicious Client}:
The malicious client makes use of a vulnerable tenant application.
The client can send requests to networked applications, including arbitrary packets.
Assuming a vulnerable application, the client can access any resource accessible by the tenant application.
While the protection of already vulnerable applications is considered out of scope,
the hosting engine must isolate other tenants' memory access from a compromised application.

\section{Architectural Design}
\label{sec:design}
\label{sec:femto-arch-design}

\begin{figure}
\centering
\begin{subfigure}{.5\columnwidth}
  \centering
  \includegraphics[width=.8\linewidth]{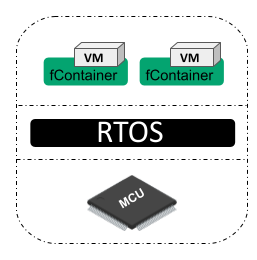}
  \caption{Femto DevOps : \\ Small VMs in containers.}
  \label{fig:sub1}
\end{subfigure}%
\begin{subfigure}{.5\columnwidth}
  \centering
  \includegraphics[width=.8\linewidth]{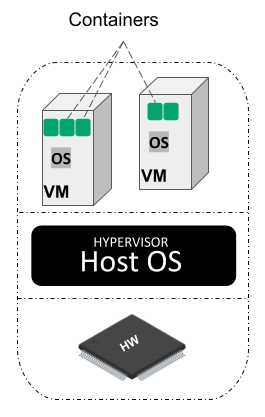}
  \caption{Cloud-native DevOps : \\ Containers in large VMs.}
  \label{fig:sub2}
\end{subfigure}
\caption{DevOps architecture: Femto \textit{vs} Cloud.}
\label{fig:arch-design}
\end{figure}

The standard system architecture enabling modern DevOps at scale in the cloud is depicted in \autoref{fig:arch-design}.2. A provider-controlled operating system hosts one or more (typically many) virtual machines sharing the same hardware resources. Each virtual machine (VM) virtualizes an OS maintained by a third-party. Each VM can host one or more (typically many) containers. This architecture is favored because it naturally provides the below properties, which are crucial in terms of code mobility and cyber-security:
\begin{enumerate}
\item \texttt{Hardware abstraction}: the VM abstracts the hardware on top of which software is running, facilitating code portability across different hardware architectures and configurations.
\item \texttt{OS abstraction}: the container further enhances code mobility by offering standardized access to OS services, which facilitates code portability across different OS. 
\item \texttt{Isolation}: the VM provides a natural defence line. If an attackers breaks out of a container, it only compromises the one VM it is in, other VMs and the host system are safe.
\end{enumerate}

However, the standard "cloud-native" DevOps architecture depicted \autoref{fig:arch-design}.2 faces issues on low-power IoT devices with constrained resources:
\begin{itemize}
\item Full OS virtualization leads to a prohibitive toll on resources and/or execution speed;
\item Standard containers solutions are not applicable on constrained IoT devices.
\end{itemize}

For these reasons, we introduce Femto-Containers, an alternative architecture targeting constrained IoT devices, as described in the following.

\subsection*{Femto-Container Architecture}
Instead of using traditional containers inside large VMs which virtualize a full-blown OS,
we host smaller VMs inside simplified containers, running on top of a real-time operating system (RTOS), as depicted in \autoref{fig:arch-design}.1.
By flipping around virtualization and containerization,
we can retain the crucial properties w.r.t. code mobility and cyber-security
(as we still combine isolation, hardware/OS abstraction)
but we are able to drastically reduce the scope of virtualization and its cost on constrained IoT devices.

Compared to the large scale cloud architecture, the Femto-Container architecture trades features for a more lightweight approach suitable for small embedded devices.
The Femto-Container architecture relies on a number of RTOS-provided features and a set of assumptions, listed below.

\subsubsection*{Use of an RTOS with Multi-Threading}
It is assumed that the RTOS supports real-time multi-threading with a scheduler.
Each Femto-Container runs in a separate thread.
Well-known operating systems in this space can provide for that, such as RIOT~\cite{baccelli2018riot} or FreeRTOS~\cite{Barry:2012} and others.
These  can run on the bulk of commodity microcontroller hardware available. 
Note that RTOS facilities for scheduling enable simple controlling of how Femto-Containers interfere with other tasks in the embedded system.

\subsubsection*{Use of Simple Containerization}
A slim environment around the \ac{VM} exposes RTOS facilities to the \ac{VM}.
The container sandboxing a VM allows this VM to be independent of the underlying operating system, 
and provide the facilities as a generic interface to the \ac{VM}.
Simple contracts between container and RTOS can be used to define and limit the privileges of a container regarding its access to OS facilities.
Note that such limitations must be enforced at run-time to safely allow 3rd party module reprogramming.

\subsubsection*{Use of Light-weight Virtualization}
The virtual machine provides hardware agnosticism, and should therefore not rely on any specific hardware features or peripherals.
This allows for running identical application code on heterogeneous hardware platforms.
The virtual machine must have a low memory footprint, both in Flash and in RAM, per VM.
This allows to run multiple VMs in parallel on the device.
Note that, since we aim to vitualize less functionalities, the VM can in fact implement a reduced feature set.
For instance, virtualized peripherals such as an interrupt controller are not required, and we give up the possibility of virtualizing a full OS.

\subsubsection*{Isolation \& Sandboxing through Virtualization}
The OS and Femto-Containers must be mutually protected from malicious code, as described in \autoref{sec:isolationmodel}.
This implies in particular that code running in the \ac{VM} must not be able to access memory regions outside of what is allowed.
Here again, simple contracts can be used to define and limit memory and peripheral access of the code running in the Femto-Container.

\subsubsection*{Event-based Launchpad Execution Model}
Femto-containers are executed on-demand, when an event in the RTOS context calls for it.
Femto-container applications are rather short-lived and have a finite execution constraint.
This execution model fits well with the characteristics of most low-power IoT software.
To simplify containarization and security-by-design, we mandate that 
Femto-Containers can only be attached and launch from predetermined launch pads, which are sprinkled throughout the RTOS firmware.
Where applicable however, the result from the Femto-Container execution can modify the control flow in the firmware as defined in the launch pad.


In the following, with a view to implement the Femto-Container architecture, we explore building blocks alternatives.

\section{Software-based Techniques for Process Isolation \& Virtualization}
\label{sec:techniques-survey}

As seen in Section \autoref{sec:femto-arch-design}, process isolation and virtualisation is a critical building block on which the femto-container architecture depends. 
Different approaches are possible in order to virtualize and/or isolate software modules running on a microcontroller.

\subsection*{Software-based Online Approach}
One type of approach to isolate software modules is to modify the embedded hardware architecture. Prominent examples of this trend include TrustZone modifying the Arm Cortex-M architecture~\cite{pinto2019trustzone}, Sanctum modifying the RISC-V architecture~\cite{sanctum-riscv}, or Sancus complementing the MSP430 architecture~\cite{alder2021aion,noorman2017sancus}. However, such \emph{hardware-based} approaches are by nature specific to each hardware architecture.
To retain more general applicability, including on legacy IoT hardware, in this paper we aim instead for a \emph{software-based} approach which does not require specific hardware-based memory protection mechanism.

Taking a different angle, software security guarantees (such as process isolation, or functional correctness) can be determined with \emph{offline} techniques such as formal verification, which can prove properties on software that is known a priori, before it is actually deployed and running. In this paper, we do not pursue this type of approach. We instead focus on \emph{online} techniques, which enforce checks and guarantees on-the-fly, on previously unknown code which is deployed (and runs tentatively). The reason for this choice is mainly that our use cases include scenarios where distinct entities indepently update different software modules running on the same IoT device.

Nevertheless, note that our approach does not preclude the complementary use of hardware-based mechanisms and/or offline mechanisms.

\subsection{Survey of Software-based Techniques}

Different categories of light-weight, software-based techniques for process isolation and virtualization have been developed in prior work.

\paragraph{Virtual machines}
One category of techniques consists in small virtual machine, used to host and isolate a process from other processes running on the microcontroller.
One example is WebAssembly (Wasm~\cite{haas2017wasm}), a \ac{VM} specification with a stack-based architecture, designed for process isolation in Web browsers, which has recently been ported to microcontrollers~\cite{wasm3}.
Another recent example is Velox~\cite{tsiftes2018velox} a VM able to host and isolate high-level functional programming logic on microcontrollers.
On the other hand, Darjeeling~\cite{darjeeling} is a subset of the Java VM, modified to use a 16 bit architecture, designed for 8- and 16-bit microcontrollers.
In fact, beyond the low-power IoT domain, small Java \acp{VM} have also been used in other contexts for a long time.
For instance JavaCard~\cite{identity2018javacard} uses a small Java \ac{VM} running on smart cards.

A different VM approach comes from the Linux ecosystem, based on eBPF (extended Berkeley Packet Filter~\cite{mccanne1993bpf,fleming2017eBPF}), which enables small \ac{VM} hosting and isolating, for debug and inspection code inserted in the Linux kernel at run-time.
Recently, a preliminary prototype adapting an eBPF virtual machine hosted on a low-power microcontroller was developed in~\cite{zandberg2020minimal}.

\paragraph{Scripted logic containers}
Yet another type of approach uses scripted logic interpreters to virtualize and/or isolate some processes.
For instance, MicroPython~\cite{micropython} is a very popular scripted logic interpreter used on microcontrollers, offering partial Python scripting support.
Another popular scripted logic interpreter is JerryScript, which offers full ECMA5.1 scripting support.
Prior work such as RIOTjs~\cite{riot-js-container} provides a small JavaScript run-time container, which can host (updatable) business logic interpreted on-board a microcontroller, using JerryScript glued atop a real-time OS (RIOT).
However, neither MicroPython nor RIOTjs/JerryScript provide specific container isolation guarantees.
Complementary mechanisms can however guarantee mutual isolation between scripts.
For instance, the SecureJS~\citep{ko2021securejs}, provides a Javascript-to-Javascript compiler (used on top of JerryScript).
However SecureJS does not target low-power microcontrollers specifically.

\paragraph{OS-level mechanisms}
Yet another category of solution uses OS-level mechanisms for process isolation.
For instance, Tock~\cite{tockos} is an OS written in the Rust programming language, which offers strong isolation between its kernel and application logic processes.
However, Tock requires that microcontroller hardware provides a \ac{MPU}.
More distant related work can also be found in the domain of network function virtualization. Compact kernels such as EdgeOS~\citep{ren2020fine} provide light-weight instance spinning and isolated execution mechanisms. However such kernels are designed for high-throughput middleboxes (which are Linux-capable) instead of low-throughput, low-power microcontrollers.

\subsection{Candidate Techniques pre-Selection}
We next aim to provide a reality check gauging the potential of the different categories of approaches. For this, we pre-select and compare a representative subset of the existing solutions we identified.

In the \emph{virtual machines} category, we selected WebAssembly~\cite{wasm3} and rBPF~\cite{zandberg2020minimal}. Our reasoning motivating our choices here is that WebAssembly is a well-known solution for strong, generic software module isolation, while rBPF promises very small memory footprint accoring to preliminary prior work~\cite{zandberg2020minimal}.

In the \emph{scripted logic containers} category, we selected MicroPython~\cite{micropython} and RIOTjs~\cite{riot-js-container}, which are good representatives of prominent high-level scripting languages on microcontrollers: Python and JavaScript, respectively. We also remark that the performance of RIOTjs is an upper bound for that of SecureJS (since in essence SecureJS adds a layer on top of RIOTjs).

We however chose not to further pursue techniques using OS-specific mechanisms, because we aim to retain generic applicability to multiple OS in this space.
We now overview the essential aspects of each preselected candidate technique, before proceeding to comparative benchmarks in the next section.

\subsubsection{Web Assemby}
\ \\
WebAssembly (Wasm~\cite{haas2017wasm}) is standardized by the World Wide Web Consortium (W3C). Initially aimed at portable web applications, Wasm has been adapted to microcontrollers.

\paragraph*{Architecture}
Wasm is a virtual instruction set architecture (ISA) with flexible intruction size.
This ISA allows for small binary size -- decreasing the time needed to transport logic over the network, and necessary memory footprint on the IoT device.
The WebAssembly \ac{VM} is stack-based, using both a stack and a flat heap for memory storage.
While heap and stack sizes are flexible, Wasm specifications mandate memory allocations in chunks of \SI{64}{\kibi\byte} (pages).

\begin{figure}[t]
  \centering
    \includegraphics[width=\columnwidth]{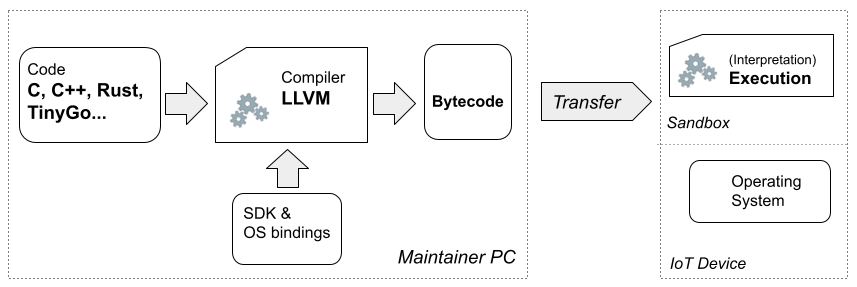}
  \caption{VM toolchain.}
  \label{fig:VM-toolchain}
    \vspace{-1.5em}
\end{figure}

\paragraph*{DevOps Toolchain}
Wasm uses the LLVM compiler: Wasm applications code can be written in any language supported by LLVM such as C, C++, Rust, TinyGo, or D, among others.
The Wasm code development and execution workflow is shown in \autoref{fig:VM-toolchain}.
Note that for C and C++, the WebAssembly binaries are created using the emcc toolchain, which combines the EmSDK with LLVM.
Furthermore, a POSIX-like interface is specified for host OS access, called WASI~\cite{WASI}. WASI standarizes access to operating system facilities such as files, network sockets, clocks and random number etc.

\paragraph*{Interpreter}
Once the Wasm binary is compiled with LLVM, the resulting bytecode can be transferred to the IoT device,
on which it is interpreted and executed, as shown in \autoref{fig:VM-toolchain}.
Several interpreters have been developed.
In this paper we use the WASM3~\cite{wasm3} interpreter.
WASM3 is based on a two-stage approach:
in a first phase, the loaded application is transpiled to an executable,
then in a second phase, it is executed in the interpreter.

\paragraph*{Security \& Isolation}
The sandbox provided by Wasm offers strong security guarantees on memory access.
The memory space accessible by the virtual machine is a virtual space mapping different real memory regions.
This prevents hazardous access to the host memory.


\subsubsection{eBPF}
\ \\
Extended Berkeley Packet Filter (eBPF~\cite{fleming2017eBPF}) is a small in-kernel \ac{VM} stemming from the Linux ecosystem, compatible with Unix-like operating systems.
eBPF provides a tiny facility able to run custom VM code, inside the kernel, hooking into various subsystems.

\paragraph*{Architecture}

eBPF is 64-bit register-based \ac{VM}, using fixed-size 64bit instructions and a small ISA.
eBPF uses a fixed-sized stack (\SI{512}{\byte}) and no heap, which limits VM memory overhead in RAM.
As a replacement for a heap, a key-value store is used for storage between invocations.
Very recently, the eBPF instruction set has been ported to microcontrollers~\cite{zandberg2020minimal}. 

\paragraph*{DevOps Toolchain}
The toolchain and workflow with eBPF is akin to that of Wasm shown in \autoref{fig:VM-toolchain}. In particular, eBPF also uses the LLVM compiler to produce bytecode, and supports WM logic written in any language supported by LLVM (C, C++, Rust, TinyGo, D...).

\paragraph*{Interpreter}
Contrary to Wasm, the bytecode produce by LLVM does not need a preliminary phase to transpile, and can directly be executed by the eBPF interpreter, on the IoT device.
Several interpreters have been developed in prior work.
In this paper, we specifically consider the rBPF interpreter~\cite{zandberg2020minimal}.

\paragraph*{Security \& Isolation}
The sandbox provided by rBPF offers security guarantees on memory access and code execution.
All memory access, including to the stack, happen via register load and store instructions and are checked against simple memory access controls.
Furthermore, there are limitations on branch and jump instruction targets.
The application has no access to the program counter via registers or instructions and a jump is always direct, and relative to current program counter.
These characteristics facilitate implementations of the necessary checks at runtime to limit access and execution and thus eliminate this attack surface.
rBPF should not be vulnerable against hazardous memory access and code execution.
However according to the authors there are no limits on the execution time granted to the application.


\subsubsection{RIOTjs}
\ \\
RIOTjs is a Javascript execution environment, integrated in the RIOT operating system.
It executes high-level logic snippets written in the Javascript language, loadable at runtime over the low-power network, on the IoT device.

\paragraph*{Architecture}
The architecture is based on a two step approach where the Javascript code loaded in the runtime container is first compiled into a compact bytecode format and then interpreted -- all of which happens directly on the IoT device.
The bytecode is a CISC-like instruction set with main focus on a compact representation.
To achieve this, the instruction set uses single instructions to cover multiple atomic tasks.

\paragraph*{Interpreter \& DevOps Toolchain}
Using a small interpreter provided by the JerryScript engine~\cite{JerryScript}, RIOTjs provides a relatively lightweight VM.
JerryScript performs a pre-execution phase which parses the Javascript code.
The parser itself is implemented as a recursive descent parser to convert the javascipt source into bytecode, without requiring an abstract syntax tree.
Thanks to this parsing phase, the toolchain is simplified: the container developer only needs a text editor (which is a major advantage of scripted logic approaches).

\paragraph*{Security \& Isolation}
With RIOTjs, compiled bytecode is executed within a virtual machine.
However, RIOTjs is not specifically designed for JavaScript runtime container security and isolation.
The hardware-specific mechanisms (e.g. hardware memory protection) or additional layers and complementary mechanisms in software (e.g. SecureJS~\citep{ko2021securejs} already mentionned earlier) would be necessary to provide strong security and isolation. guarantees.

An alternative which one can envision is to offload parsing and bytecode generation to the maintainer PC. However, in this case, an additional layer providing mechanisms that check bytecode correctness would be required on the IoT device.

Hence, the benchmarks results we provide in the next section are to be considered as an upper bound on what can be achieved with this type of approach.

\subsubsection{MicroPython}
\ \\
MicroPython~\cite{micropython} is akin to the JerryScript engine, but for Python code.
Bare-metal operation of MicroPython is possible on some IoT hardware such as the pyboard.
Variants of MicroPython such as CircuitPython~\cite{circuitpython} can run bare-metal on more hardware.
Integration in operating systems is also available (such as in ~\cite{RIOT-micropython}, which is similar to RIOTjs, but for MicroPython).

\paragraph*{Architecture}
MicroPython is based on a two stage approach.
In a first phase, based on an abstract syntax tree, a parser/lexer compiles Python code to native bytecode.
In a second phase, an interpreter executes the bytecode.
Both stages can be performed directly on the IoT device.
Within the Python logic hosted in the container, memory management is abstracted away for the developer: it automated with a heap and garbage collector.

\paragraph*{Interpreter \& DevOps Toolchain}
The bytecode interpreter machine implements a stack-based architecture.
Thanks to the parsing phase which can be performed by MicroPython, a minimal toolchain is possible: the container developer only requires a text editor.


\paragraph*{Security \& Isolation}
MicroPython is not specifically designed for Python runtime container security and isolation.
Additional layers and complementary mechanisms would be necessary to provide strong security and isolation guarantees.
Hence, the benchmarks results we provide in the next section are to be considered as an upper bound on what can be achieved with this type of approach.

\section{Benchmarks of Process Virtualization \& Isolation Techniques}

In this section, to get an idea of what to expects in terms of ballpark performance, we carry out benchmarks comparing the candidate virtualization and isolation techniques we pre-selected.
Based on these results, we aim to discuss and select an approach upon which to base our Femto-Containers design.

\subsection{Hardware \& Software Setup}
\begin{figure}[t]
  \centering
    \includegraphics[width=0.6\columnwidth]{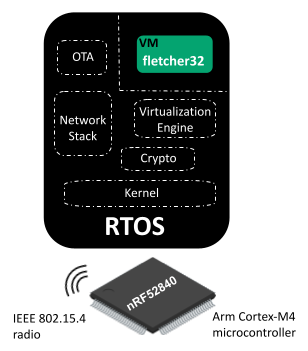}
  \caption{Benchmark setup. VM executing Fletcher32 logic, hosted in a typical RTOS configuration, on an nRF52840 microcontroller.}
  \label{fig:fletcher32-VM}
\end{figure}

The hardware and software setup is depicted in Fig.~\ref{fig:fletcher32-VM}. For each virtualization candidate, the virtual machine is loaded with an application performing a Fletcher32 checksum on a \SI{360}{\byte} input string.
All benchmarks are run COTS hardware: a Nordic nrf52840dk development kit, which is based on a common ARM Cortex-M4 processor running at \SI{64}{\mega\hertz}.
The operating system hosting the VMs is RIOT~\cite{baccelli2018riot}.
As base, we take RIOT Release 2021.04, configured to be IoT-ready, providing standard low-power networking connectivity, leveraging the board's IEEE 802.15.4 radio chip and a ressource-efficient IPv6-compliant stack (6LoWPAN, UDP, CoAP and SUIT).

\subsection{Results \& Preliminary Analysis}

\begin{table}[ht]
  \centering
  \begin{tabular}{lrr}
    \toprule
    & ROM size & RAM size \\
    \midrule
    WASM3 Interpreter  & \SI{64}{\kibi\byte} & \SI{85}{\kibi\byte} \\
    rBPF Interpreter   & \SI{4,4}{\kibi\byte} & \SI{0,6}{\kibi\byte} \\
    RIOTjs        & \SI{121}{\kibi\byte} & \SI{18}{\kibi\byte} \\
    MicroPython        & \SI{101}{\kibi\byte} & \SI{8,2}{\kibi\byte} \\
    \midrule
    Host OS (without VM)   & \SI{52,5}{\kibi\byte} & \SI{16,3}{\kibi\byte} \\

    \bottomrule
  \end{tabular}
  \vspace{0.5em}
\caption{Memory requirements for VM interpreters.}
\label{tbl:vmsize}
  \vspace{-1.5em}
\end{table}

\begin{table}[ht]
  \centering
  \begin{tabular}{lrrr}
    \toprule
    & code size & startup time &run time \\
    \midrule
    Native C         & \SI{74}{\byte}  & -- &\SI{27}{\micro\second} \\
    WASM3            & \SI{322}{\byte} & \SI{17096}{\micro\second} &\SI{980}{\micro\second} \\
    rBPF             & \SI{456}{\byte} & \SI{1}{\micro\second} & \SI{2133}{\micro\second} \\
    RIOTjs      & \SI{593}{\byte} & \SI{5589}{\micro\second} & \SI{14726}{\micro\second} \\
    MicroPython      & \SI{497}{\byte} & \SI{21907}{\micro\second} & \SI{16325}{\micro\second} \\
    \bottomrule
  \end{tabular}
  \vspace{0.5em}
  \caption{Size and performance of fletcher32 logic hosted in different VMs.}
  \label{tbl:scriptsize}
  \vspace{-1.5em}
\end{table}

Our benchmarks results comparing different scripting and virtualization techniques are shown in Table~\ref{tbl:vmsize} and Table~\ref{tbl:scriptsize}.
The results highlight how much the footprint of hosting logic in a VM can vary, depending on the virtualization technique being used.

\paragraph*{Looking at size}
While the size of applications are roughly comparable accross virtualization techniques (see Table~\ref{tbl:scriptsize}) the memory required on the IoT device differs wildly.
In particular, techniques based on script interpreters (RIOTjs and MicroPython) require the biggest dedicated ROM memory budget, above \SI{100}{\kibi\byte}.

For comparison, the biggest ROM budget we measured requires 27 times more memory than the smallest budget.
Similarly, RAM requirements vary a lot. Note that we could not determine with absolute precision the lower bound for script interpreters techniques (due to some flexibility given at compile time to set heap size in RAM). Nevertheless, our experiments show that the biggest RAM budget requires 140 times more RAM than the smallest budget.
We remark that, as noted in prior work~\cite{zandberg2020minimal} the minimum required page size of \SI{64}{\kibi\byte}  to comply with the WebAssembly specification explains why WASM performs poorly in terms of RAM. One can envision enhancements where this requirement is relxaed. However the RAM budget would still be well above an order of magnitude more than the lowest RAM budget we measured (rBPF).

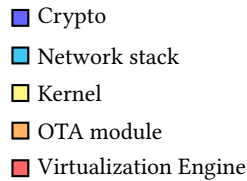
\begin{figure*}[t]

\begin{subfigure}{\columnwidth}
    \begin{tikzpicture}[scale=0.7]
 \pie[explode=0.1, text=inside, radius=2.5]
    {14/,
     38/,
     32/,
     16/
     }
    \end{tikzpicture}
    \caption{RIOT without hosting engine(53kBytes).}
  \end{subfigure}
   \begin{subfigure}{\columnwidth}
    \begin{tikzpicture}[scale=0.7]
 \pie[explode=0.1, text=inside, radius=2.5]
    {5/,
     13/,
     11/,
     5/,
     66/
     }
    \end{tikzpicture}
    \caption{RIOT with MicroPython VM (154kBytes).}
  \end{subfigure}
  \begin{subfigure}{\columnwidth}
    \begin{tikzpicture}[scale=0.7]
 \pie[explode=0.1, text=legend, radius=2.5]
    {13/Crypto,
     35/Network stack,
     30/Kernel,
     14/OTA module,
     8/Virtualization Engine
     }
    \end{tikzpicture}
    \caption{RIOT with rBPF virtual machine (57kBytes).}
  \end{subfigure}
  \caption{Flash memory distribution. RIOT with 6LoWPAN, CoAP, SUIT-compliant OTA and different application hosting engines.}
  \label{fig:candidates-memory-distrib}
\end{figure*}

Last but not least, let's give some perspective by comparison with a typical memory budget for the \emph{whole} software embedded on the IoT device.  As a reminder, in the class of devices we consider, a microcontroller memory capacity of 64kB in RAM and 256kB in Flash (ROM) is not uncommon. A typical OS footprint for this type of device is shown in the last row of Table~\ref{tbl:vmsize}. For such targets, according to our measurements, adding a VM can either incur a tremendous increase in total memory requirements (200\% more ROM with MicroPython) or a negligible impact (8\% more ROM with rBPF) as visualized in \autoref{fig:candidates-memory-distrib}.

\paragraph*{Looking at speed}
To no surprise, beyond size overhead, virtualization also has a cost in terms of execution speed.
But here again, performance varies wildly depending on the virtualization technique.
On one hand, solutions such as MicroPython and RIOTjs directly interpret the code snippet and execute it.
On the other hand, solutions such as rBPF and WASM3 require a compilation step in between to convert from human readable code to machine readable.

Our measurements show that script interpreters incur an enormous penalty in execution speed.
Compared to native code execution, script interpreters are a whooping ~600 times slower.
Compared to the same base (native execution) WASM is only 37 times slower, and rBPF 77 times slower.

One last aspect to consider is the startup time dedicated to preliminary pre-processing when loading new VM logic, before it can be executed (including steps such as code parsing and intermediate translation, various pre-flight checks etc.).
Depending on the virtualization technique, this startup time varies almost 1000 fold -- from a few microseconds compared to a few milliseconds.

\subsection{Discussion}

We now aim to reason a choice for an approach to design tiny \acp{VM} which efficiently address our target use cases (described in Section~\ref{sec:case-study}) which involve hosting and mutually isolating multiple VMs which may contain either high-level business logic, or low-level debug/monitoring code snippets.

\paragraph{Considering architecture \& security}
There are notable architectural differences amongst the solutions we pre-selected and looked at in this this section.
For instance, WASM, MicroPython and RIOTjs each require some form of heap on which to allocate application variables.
On the other hand, rBPF does not require a heap.
With a view to accommodating several VMs concurrently, a heap-based architecture presents on the one hand some potential advantages in terms of memory (pooling) efficiency, but on the other hand some potential drawbacks in terms of security (mututal isolation of the VMs' memory).

From another angle: security guarantees call for a formally verified implementation of the hosting engine, down the road.
A typical approximation is: less LoC (lines of code) means less effort produce a verified implementation.
For instance, the rBPF implementation is ~1,5k LoC, while the WASM3 implementation is ~10k LoC. The other implementations we considered in our pre-selection (RIOTjs and MicroPython) encompass significantly more LoC.

\paragraph{Considering performance}
Our benchmarks indicate that both from a memory overhead and from a startup time standpoint an eBPF-based approach is the most attractive, by far.
From an execution time point of view however, a WebAssembly approach does offer faster execution times than an eBPF-based approach. We nevetherless deem safe to consider that, for the use cases we target, this difference is negligible.
These results both extend and confirm independent results presented in prior work~\cite{zandberg2020minimal}.
All in all, both due to much larger memory footprint and enormous execution time penalties, Python and JavaScript approaches could not be considered beyond rapid prototyping -- in particular when considering one of our uses cases: a virtual machine hosting debug applications in a hot code path.

\paragraph{Ahead-of-Time vs Just-in-Time}
One approach to speed up embedded execution time is to perform a translation into device-native code.
One way to offload the device is to use more Ahead-of-Time (AoT) compilation/interpretation, and less Just-in-Time (JiT) processing on-device. 
However, using AoT pre-compiled code can both complicate run-time security checks on-board the IoT device, and reduce the portability of the code deployed on the device.
For these reasons, in this paper, we consider primarily JiT.

\paragraph{Intermediate conclusion}
Based on our discussion and our benchmarks, we derive the preliminary conclusion that an eBPF-based architecture is a promising approach to design efficient and secure tiny concurrent containers to host and execute logic on a microcontroller.
In the following, we thus design, implement and experimentally evaluate a novel virtualization and isolation mechanism for software modules, based on eBPF.

\section{Femto-Container Implementation}
As proof of concept, we implemented the femto-container architecture, with containers hosted in the operating system RIOT and virtualization using an instruction set compatible with the eBPF instruction set.
This implementation is open source (published in \cite{femto-github}). We detail below its main caracteristics.

\if{0}
The Femto-Container engine is designed to host and interpret one or more small virtual machines, hosted on top of an OS, running on a microcontroller-based IoT device.
The virtual machine instruction set is compatible with the eBPF instruction set architecture.
The address space inside the virtual machine is shared with the operating system, no memory address translation is used by default.
\fi

\subsection*{Use of RIOT Multi-Threading}
\begin{figure}[t]
  \centering
    \includegraphics[width=\columnwidth]{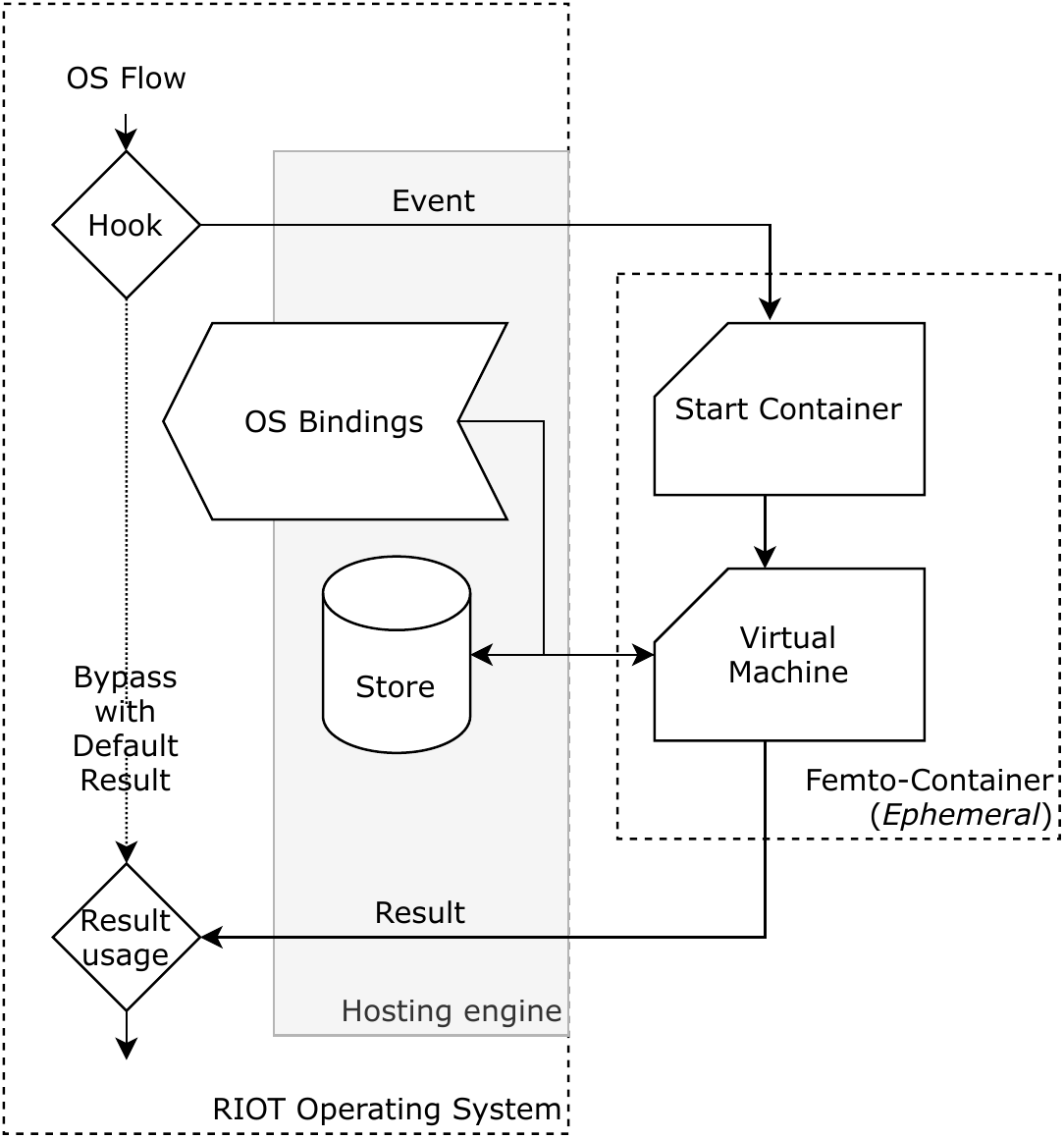}
  \caption{Femto-Container RTOS integration.}
  \label{fig:vm-overview}
\end{figure}

Each Femto-Container application instance running on the RTOS is scheduled as a regular thread in RIOT.
The native OS thread scheduling mechanism can thus simply execute concurrently  and share resources amongst multiple femto-containers and other tasks, spread over different threads.
An overview of how Femto-Containers integrates in the operating system is shown in \autoref{fig:vm-overview}.
A femto-container instance requires minimal RAM: a small stack and the register set, but no heap.
The host RTOS bears thus a very small overhead per femto-container instance.

The  hardware and peripherals available on the device are not accessible by the Femto-Container instances.
All interaction with hardware peripherals passes through the host RTOS via the system call interface.
As the Femto-Container virtual machine does not virtualize its own set of peripherals, no interrupts or pseudo-hardware is available to the Femto-Container application. This also removes the option to interrupt the application flow inside a Femto-Container.

\subsection*{Basic Containerization}
Simplified containers provides a uniform environment around the VM, independent of the operating system (RIOT in our implementation).
Access from the femto-container to the required OS facilities is allowed through system calls to services provided by RIOT.
These system calls can be used by the loaded applications via the eBPF native \texttt{call} instruction.
Furthermore, the OS can share specific memory regions with the container.
\ \\
{\bf Key-value store.}
In lieu of a file system, applications hosted in femto-containers can load and store simple values, by a numerical key reference, in a key-value store.
This provides a mechanism for persistent storage, between application invocations.
Interaction with this key-value store is implemented via a set of system calls, keeping it independent of the instruction set.
By default, two key-value stores are provided by the OS.
The first key-value store is local to the application, for values that are private to the VM accommodated in the container.
The second key-value store is global, and can be accessed by all applications, used to communicate values between applications.
An optional third intermediate-level of key-value store is possible to facilitate sharing data across a set of VMs from the same tenant, while isolating this set of VMs from other tenants' VMs.

\subsection*{Light-weight Virtualisation}
Application code is virtualized using the eBPF instruction set and the rBPF interpreter.
\ \\
{\bf Register-based VM.} The virtual machine operates on eleven registers of 64 bits wide.
The last register (\texttt{r10}) is a read-only pointer to the beginning of a \SI{512}{\byte} stack provided by the femto-container hosting engine.
Interaction with the stack happens via load and store instructions.
Instructions are divided into an \SI{8}{\bit} opcode, two \SI{4}{\bit} registers: source and destination, an \SI{16}{\bit} offset field and an \SI{32}{\bit} immediate value.
Position-independent code is achieved by using the reference in \texttt{r10} and the offset field in the instructions.

{\bf Jumptable \& Interpreter.}
The interpreter parses instructions and executes them operating on the registers and stack.
The machine itself is implemented as a computed jumptable, with the instruction opcodes as keys.
During execution, the hosting engine iterates over the instruction opcodes in the application,
and jumps directly to the instruction-specific code.
This design keeps the interpreter itself small and fast.

\subsection*{Isolation \& Sandboxing}
To control the capabilities of Femto-Containers, and to protect the OS from memory access by malicious applications, a simple but effective memory protection system is used.
By default each virtual machine instance only has access to its \ac{VM}-specific registers and its stack.

\begin{figure}[t]
  \centering
    \includegraphics[width=\columnwidth]{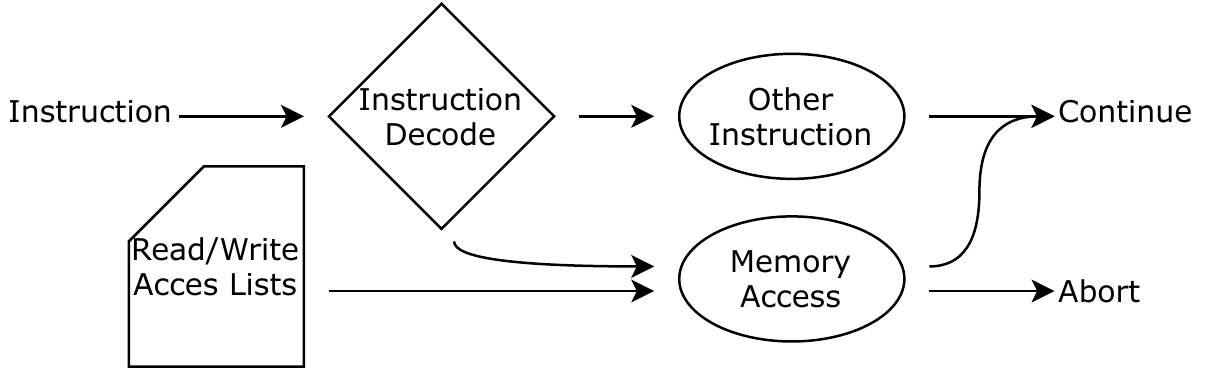}
  \caption{Interaction between memory instructions and the access lists.}
  \label{fig:memaccess}
\end{figure}

{\bf Memory access checks at runtime.}
Whitelist can be configured (attached in the hosting engine) to explicitly allow a \ac{VM} instance access to other memory regions.
These memory regions can have individual flags for allowing read/write access.
For example, a firewall-type trigger can grant read-only access to the network packet, allowing the virtual machine to inspect the packet, but not to modify it.
As the memory instructions allow for calculated addresses based on register values, memory accesses are checked at runtime against the resulting address, as show in \autoref{fig:memaccess}. Illegal access aborts execution.

{\bf Pre-flight instruction checks.}
A femto-container verifies the application before it is executed for the first time.
These checks includes checks on the instruction fields.
For example, as there are only 11 registers, but space in the instruction for 16 registers, the register fields must be checked for out-of-bounds values.
A special case here is register \texttt{r10} which is read-only, and thus is not allowed in the destination field of the instructions.

The jump instructions are also checked to ensure that the destination of the jump is within the address space of the application code.
As calculated jump destinations are not supported in the instruction set, the jump targets are known before executions and are checked during the pre-flight checks.
During the execution of the application, the jump destinations no longer have to be verified and can be accepted as valid destinations.

Finite execution is also enforced, by limiting both the total number of instructions $ N_i $, and the number of branch instructions $ N_b $ that are allowed.
In practice, this limits the total number of instructions executed to:
$ N_i \times N_b $.

\subsection*{Hooks \& Event-based Execution}
The Femto-container hosting engine instantiates and runs containers as triggered by events within the RTOS.
Such events can be a network packet reception, sensor reading input or an operating system scheduling events for instance.
Business logic applications can be implemented either by directly responding to sensor input or by attaching to a timer-based hook to fire periodically.

Simple hooks are pre-compiled into the RTOS firmware, providing a pre-determined set of pads from which femto-containers can be attached and launched.

\begin{lstlisting}[language=c,label=lst:hook,caption={Example hook implementation.}]
sched_ctx_t context = {
    .previous = active_thread,
    .next = next_thread,
};

int64_t result;

fc_hook_execute(BPF_HOOK_SCHED, &context,
                sizeof(context), &result);
\end{lstlisting}

An example of a hook integrated in the firmware is shown in \autoref{lst:hook}.
The firmware has to set up the context struct for the Femto-Containers after which it can call the hosting engine to execute the containers associated with the hook.


\subsection*{Low-power Secure DevOps}

Launching a new femto-container or modifying an existing femto container can be done without modifying the RTOS firmware. However, updating the hooks themselves requires a firmware update.
In our implementation, both types of updates use CoAP network transfer and software update metadata defined by SUIT~\cite{moran2019firmware} (CBOR, COSE) to secure updates end-to-end over network paths including low-power wireless segments~\cite{zandberg2019secure}.
Leveraging SUIT for these update payloads provides authentication, integrity checks and roll-back options.
Updating a Femto-Container application attached to a hook is done via a SUIT manifest.
The exact hook to attach the new Femto-Container to is done by specifying the hook as unique identifier (UUID) as storage location in the SUIT manifest.
A rapid develop-and-deploy cycle only requires a new SUIT manifest with the storage location specified every update.
Sending this manifest to the device triggers the update of the hook after the new Femto-Container application is downloaded to the device and stored in the RAM.
Multiple Femto-Containers can be deployed on a single hook, where it depends on the hook implementation how conflicting return values from the different Femto-Containers are used.
For example, a timer-based hook for periodic container execution support attaching multiple Femto-Containers from different tenants.
This hook then periodically executes all containers attached without using any of the return values.

\section{Use-Case Prototyping with Femto-Containers}
\label{sec:femto-container-examples}
In this section, we use Femto-Containers to prototype the implemention of several use cases involving one or more applications, hosted concurrently on a microcontroller, matching targets we identified initially (in \autoref{sec:case-study}).
In the prototype implementation we show below, we used C to code logic hosted in Femto-Containers.
However, any other language compiled with LLVM could be used instead (C++, Rust, TinyGo, D...).


\subsection{Kernel Debug Code Example}
\label{sec:femto-debug-example}
The first prototype consists in a single application, which intervenes on a hot code path:
it is invoked by the scheduler of the OS, to keep an updated count of threads' activations.
The logic hosted in the Femto-Container is shown in \autoref{lst:thread_log}.
A small struct is passed as context, which contains the previous running thread ID and the next running thread ID.
The application maintains a value for every thread, incrementing it every time the thread is scheduled.
External code can request these counters and provide debug feedback to the developer.

\begin{lstlisting}[language=c,label=lst:thread_log,caption={Thread counter code.}]
#include <stdint.h>
#include "bpf/bpfapi/helpers.h"

#define THREAD_START_KEY    0x0

typedef struct {
    uint64_t previous; /* previous thread */
    uint64_t next;     /* next thread */
} sched_ctx_t;

int pid_log(sched_ctx_t *ctx)
{
 /* Zero pid means no next thread */
    if (ctx->next != 0) {
        uint32_t counter;
	uint32_t thread_key = THREAD_START_KEY +
                ctx->next;
        bpf_fetch_global(thread_key,
                         &counter);
        counter++;
        bpf_store_global(thread_key,
                         counter);
    }
    return  0;
}
\end{lstlisting}

\subsection{Networked Sensor Code Example}
\label{sec:femto-sensor-example}
\begin{figure}[t]
  \centering
    \includegraphics[width=0.8\columnwidth]{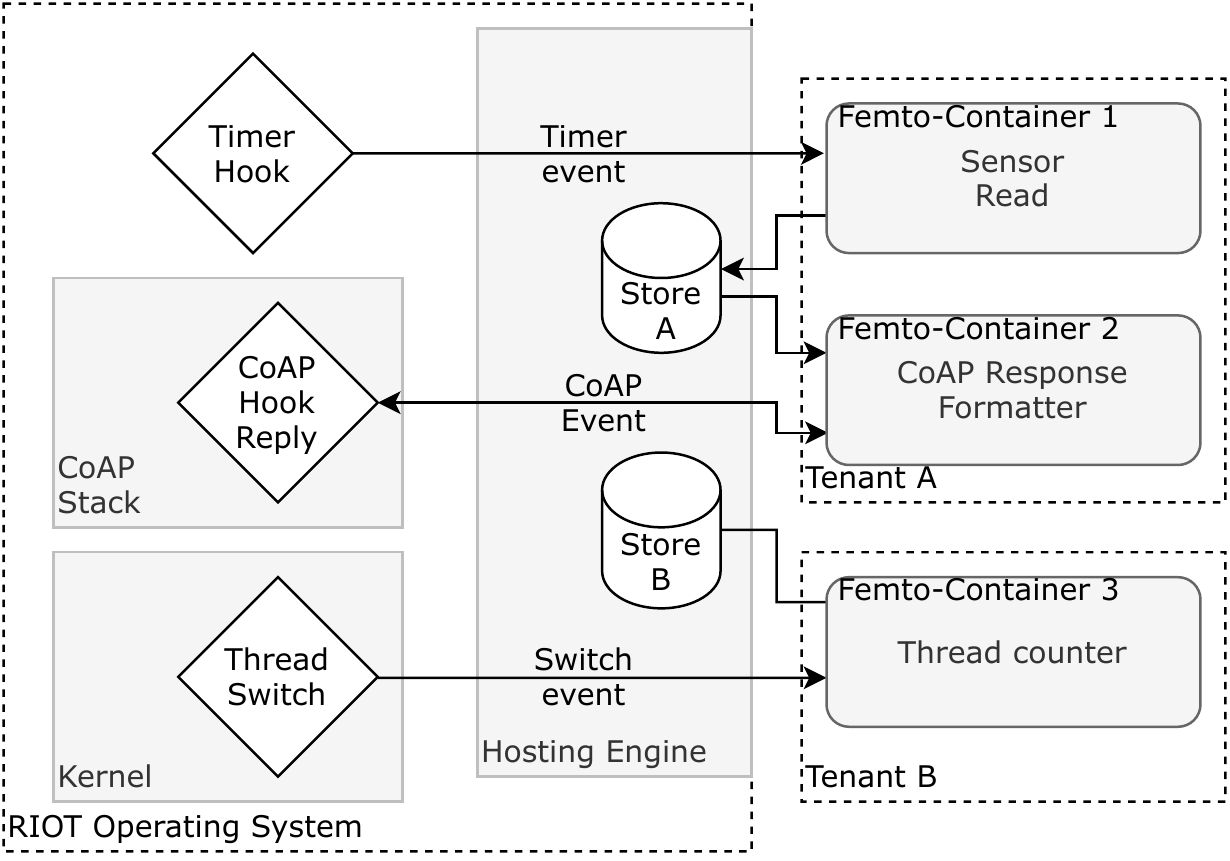}
  \caption{Event and value flow when hosting multiple containers for different tenants.}
  \label{fig:inter-vm}
\end{figure}

For the second prototype we add two Femto-Containers from another tenant to the setup of the first prototype.
Interaction between these two additional containers is achieved via a separate key-value store, as depicted in \autoref{fig:inter-vm}.
The logic hosted in the first Femto-Container, periodically triggered by the timer event, reads, processes and stores a sensor value.
The code for this logic is shown in \cite{femto-sensor-process}.
The second container's logic is triggered upon receiving a network packet (CoAP request), and returns the stored sensor value back to the requestor.
The code for this logic is shown in \cite{femto-fetch}.

In this toy example, the sensor value processing is a simple moving average, but more complex post-processing is possible instead, such as differential privacy or some federated learning logic, for instance.
This example sketches both how multiple tenants can be accommodated, and how separating the concerns between different containers is possible (between sensor value reading/processing on the one hand, and on the other hand the communication between the device and a remote requestor).

\section{Performance Evaluation}

In this section we evaluate the performance of Femto-Containers on low-power IoT hardware.
We use RIOT Release 2021.04 as a base for our benchmarks.

\subsection{Hardware Testbed Setup}
We carry out our measurements on popular, commercial, off-the-shelf IoT hardware, representative of the landscape of modern 32-bit microcontroller architecture that are available: Arm Cortex-M, ESP32, and RISC-V. More precisely, we build and run the code on the following boards:
\begin{itemize}
    \item a Nordic nRF52840 Development Kit, using an Arm Cortex-M4 microcontroller
        with \SI{256}{\kibi\byte} RAM,
        \SI{1}{\mebi\byte} Flash,
        and a \SI{2.4}{\giga\hertz} radio transceiver (BLE/802.15.4)
    \item a WROOM-32 board, using an ESP32 module 
        which provides two low-power
        Xtensa\textsuperscript{\tiny\textregistered} 32-bit LX6
        microprocessors with integrated Wi-Fi and Bluetooth,
        \SI{520}{\kibi\byte} RAM,
        \SI{448}{\kibi\byte} ROM
        and \SI{16}{\kilo\byte} RTC SRAM.
    \item a Sipeed Longan Nano GD32VF103CBT6 Development Board,
        which provides a RISC-V 32-bit microcontroller
        with \SI{32}{\kibi\byte} RAM
        and \SI{128}{\kibi\byte} Flash.
\end{itemize}
An open-access testbed such as IoT-Lab~\cite{iot-lab2015-ieee-iot-wf} also provides some of this hardware, for reproducibility.


\subsection{Femto-Container Engine Code Analysis}

The Femto-Containers hosting engine code size is small: 1874 lines of code in total.
This includes bindings to the operating system facilities.
Compared to the rBPF hosting engine for example (1615 lines of code)
this represents a relatively modest increase (15\%) which remains in the same ballpark.

The in-memory structures required to run Femto-Containers are also small. 
There are two important structures used to manage Femto-Containers.
The first structure contains the full state of the virtual machine and any flags required to manage the \ac{VM}.
This structure requires \SI{664}{\byte} in total and includes the stack for the \ac{VM} instance.
The second structure is a small \SI{16}{\byte} structure used to whitelist different memory regions for additional access for the \ac{VM}.
The virtual machine state structure already includes such a memory whitelist structure to grant access to the stack space of the \ac{VM}.

\subsection{Experiments with a Single Container}

In this section, we evaluate the footprint and the speed of execution with a Femto-Container, on various 32-bit microcontrollers.
Again, we compare to the performance of using a basic rBPF virtual machine.
Femto-Containers proceeds to security checks on the application bytecode, prior to actually launching the VM.
First, this verification stage checks that the registers in all instructions are within the bounds of the eleven available registers, where the source address must be one of these registers and the destination must be one of register \texttt{r0} to \texttt{r9}, as register \texttt{r10} is read-only per specification.
Second, the verification checks the destination of the branch-type instructions.
Compared to rBPF which performs more limited checks, for instance, the Femto-Container engine increases security measures.
We evaluate the impact of these pre-flight checks on speed in \autoref{tbl:speed_comparison}.
While still much faster than alternative virtualization techniques (recall \autoref{tbl:scriptsize}), Femto-Containers startup time is significantly longer than those in rBPF, across the board.
Note however that pre-flight checks are run only the first time the VM runs, and are skipped onwards, after the second time the application is run.
The increased security measures (in the pre-flight checks) also show when looking at the memory footprint of the Femto-Containers, reported in \autoref{tbl:femtosize}. Compared to rBPF the required ROM for the hosting engine increased slightly (less than 10\%).
The required RAM per virtual machine also increased slightly because of the additional security features.
In \autoref{tbl:app_comparison}, we also measure the RAM memory required to run the kernel debug application example (described in \autoref{sec:femto-debug-example}) which is also small, at ~700Bytes.
Last but not least, we observe in \autoref{tbl:speed_comparison} that the execution time with a Femto-Container is slighlty faster than execution time with rBPF virtual machine.

\begin{table}[ht]
  \centering
  \begin{tabular}{lrrrr}
    \toprule
    & \multicolumn{2}{c}{rBPF} & \multicolumn{2}{c}{Femto-Containers} \\
    & Install & Run & Install & Run\\
    \midrule
    Cortex-M4  & \SI{1} {\micro\second} & \SI{2133}{\micro\second} & \SI{28}{\micro\second} & \SI{2061}{\micro\second} \\
    ESP32      & \SI{15}{\micro\second} & \SI{1817}{\micro\second} & \SI{66}{\micro\second} & \SI{1770}{\micro\second} \\
    RISC-V     & \SI{1} {\micro\second} & \SI{1248}{\micro\second} & \SI{17}{\micro\second} & \SI{1238}{\micro\second} \\
    \bottomrule
  \end{tabular}
  \vspace{0.5em}
  \caption{Speed of a Femto-Container (hosting fletcher32 logic).}
  \label{tbl:speed_comparison}
  \vspace{-1.5em}
\end{table}


\begin{table}[ht]
  \centering
  \begin{tabular}{lrr}
    \toprule
    & ROM size & RAM size \\
    \midrule
    Femto-Container   & \SI{4742}{\byte} & \SI{664}{\byte} \\ 
    rBPF Interpreter   & \SI{4440}{\byte} & \SI{660}{\byte} \\
    \bottomrule
  \end{tabular}
  \vspace{0.5em}
\caption{Memory footprint of a Femto-Container hosting fletcher32 logic (on Arm Cortex-M4).}
\label{tbl:femtosize}
  \vspace{-1.5em}
\end{table}

\subsection{Experiments with Multiple Containers}


We now measure in \autoref{tbl:app_comparison} the memory required to concurrently host multiple containers from multiple tenants on the same microcontroller, from the examples we described in \autoref{sec:femto-container-examples}.
In general, each Femto-Container needs memory to
\begin{itemize}
\item store the application bytecode
\item handle the virtual machine state and stack
\end{itemize}
This minimal default memory footprint used by a Femto-Container amounts to \SI{664}{\byte}, which is for storing the VM stack,  housekeeping structs, information about memory regions etc.
Note that on top of this basic footprint, a small additional overhead is necessary to code memory permissions.
For example, the CoAP handler container (see \autoref{fig:inter-vm}) requires additional read/write permissions to two memory regions to handle the CoAP packet, which increases the overhead by \SI{16}{\byte} per region.
Furthermore, the key-value stores are also in RAM. In this case the total RAM used by the key value stores (and houskeeping) for different tenants was \SI{340}{\byte}.
Hence, the required RAM memory we measured so as to run the example with 3 containers and 2 tenants is \SI{3.2}{\kibi\byte}.

Beyond these examples, if we consider more containers hosting larger applications (e.g., $\approx$2000Bytes) an Arm Cortex-M4 microcontroller with \SI{256}{\kibi\byte} RAM, the density of containers achievable would be of $\approx$100 instances, next to running the OS.
Femto-Containers thus allow an almost arbitrarily high density of VMs, even on small microcontrollers.

\begin{table}[ht]
  \centering
  \begin{tabular}{lrrr}
    \toprule
     & Bytecode & Container RAM & Total RAM \\
    \midrule
    Thread Counter & \SI{104}{\byte} & \SI{664}{\byte} & \SI{768}{\byte} \\
    Sensor Reader  & \SI{496}{\byte} & \SI{664}{\byte} & \SI{1160}{\byte} \\
    CoAP Handler   & \SI{264}{\byte} & \SI{696}{\byte} & \SI{960}{\byte} \\
    \bottomrule
  \end{tabular}
  \vspace{0.5em}
  \caption{RAM required to host multiple concurrent Femto-Containers applications.}
  \label{tbl:app_comparison}
  \vspace{-1.5em}
\end{table}

\subsection{Overhead Added by Hooks}

One key question is how performance is affected by preprovisionning launchpads (hooks) in the RTOS firmware.
We measure in \autoref{tbl:hook_comparison} the overhead caused by adding a hook to the RTOS workflow.
This overhead amounts to $\approx$100 clock ticks on all the hardware we tested.
Compared to the number of cycles needed for an average task in the operating system, this impact is low.
Furthermore, this overhead is less than 10\% of the number of cycles needed to execute the logic hosted in a Femto-Container.
From this observation, we can conclude that, even if this hook is on a very hot code path (as for the Thread Counter example) the performance loss is tolerable.
Conversely, the perspective of adding many hooks sprinkled in many places in the RTOS firmware is realistic without incurring significant performance loss.

\begin{table}[ht]
  \centering
	\begin{tabular}{lrr}
		\toprule
		& Empty Hook & Hook with Application \\
		\midrule
		Cortex-M4 & 109 & 1750 \\
		ESP32 & 83 & 1163 \\
		RISC-V & 106 & 754 \\
		\bottomrule
  \end{tabular}
  \vspace{0.5em}
  \caption{Hook overhead in clock ticks for the thread switch example}
  \label{tbl:hook_comparison}
  \vspace{-1.5em}
\end{table}

\section{Discussion}

\paragraph{Virtualization \emph{vs} Power-Efficiency}
Inherently, virtualization causes some execution overhead, due to interpretation of the code.
Thus Femto-Containers increase power consumption for functionality executed within the \ac{VM}, compared to native code execution.
However, this drawback is mitigated by several other factors.
First, the absolute power consumption overhead may be neglible, e.g. if the hosted logic is not performing long-lasting, heavy-duty tasks.
Second, network transfer costs, power consumption and downtime are saved if software updates modify a Femto-Container instead of the full firmware.

\paragraph{Controlling Tenant Priviledges}
Controlling and granting access to specific RTOS resources to different containers or tenants is a complex challenge.
Our design includes a basic permission system based on preprovisionned hooks, system calls, and simple contracts between the hosting engine (on behalf of the OS) and a given container.
Basically: the OS restricts the set of priviledges that can be granted, the container specifies the set of priviledges it requires, and the hosting engine grants the intersection of these sets.
One limitation of our current simplified design is that there is only one fixed set of priviledges possible per hook.
In case 2 tenants have different priviledges, a second hook must be made available.
Additional mechanisms would be required to overcome this limiation and/or to enable dynamic priviledge levels.

\paragraph{Install Time \emph{vs} Execution Time}
As mentioned before, one limitation due to virtualization is the inherent slump in execution speed, compared to native code exection.
One way to remove this overhead is to transpile the portable eBPF bytecode into native instruction code.
This could be done in a single pass to convert the whole application into native instructions in an installation step.
This can result into a speed-up at the cost of extra install-time overhead.
To avoid the issues describe before on complicating the run-time security checks, this compilation into native code has to be done at run-time by the device deploying the code.

\paragraph{Fixed- \emph{vs} Variable-length Instructions}
Originally, eBPF scripts are optimized for fast execution on 64-bit platforms.
Compared to other virtual machines such as Wasm, the resulting bytecode is relative large.
In fact, most of the instructions have bit fields that are fixed at zero.
A possible way to reduce the size of these scripts is to compress the instructions into a variable size instruction set, removing these fields from the instructions where possible.
This would create a variable length instruction set based on the eBPF set.
For example the immediate field is not used with half of the instructions and would reduce the instructions to 32 bits in size when removed.

\paragraph{Formal Verification Perspectives}
With such a small size in terms of LoC, there is a clear opportunity to go further and attempt at producing a formally verified implementation of the Femto-Containers hosting engine, providing proof that the memory of the host system and of other tenants are protected against malicious tenants and clients.
So far, we have been using fuzzing tools on the application loaded into the virtual machine in an to find bugs and vulnerabilities in the hosting engine.
However, this does not ensure that all opcodes are executed according to their specification, it only checks whether instructions are able to crash the running system.

\paragraph{rBPF Logic \& Execution Limitations}
Inherent limitations due to the eBPF instruction set, combined with the absence of virtualized hardware, restrict what logic can be deployed in Femto-Containers currently.
Femto-Containers are designed to host logic that is rather script-like, short-lived, and not computation-intensive. 
On the one hand, such characteristics increase security-by-design.
On the other hand they reduce flexibility.
For instance, asynchronous operation is not supported: there is no option to interrupt the control flow inside a Femto-Container from outside the virtual machine.
Another limitation is the fixed, small size of the stack (512 Bytes) dictated by the eBPF specification.
More memory-consuming tasks would need special handling to provide additional memory.
Allowing the application to request more stack from the RTOS, for example via the contracts, could solve part of this issue.
More computation- and memory-intensive tasks could also make use of additional system calls provided by the RTOS, which could execute generic primitives at native speed.

\section{Related work}
We have already provided a survey of related work in~\autoref{sec:techniques-survey}.
To the best of our knowledge, the closest related work is rBPF~\cite{zandberg2020minimal}. Compared to rBPF:
\begin{itemize}
\item Femto-Containers are applicable not only to single container use-cases, but also to use cases hosting multiple containers and distinct tenants, concurrently, on the same microcontroller;
\item Femto-Containers provide additional containerization, security and isolation mechanisms;
\item Femto-Containers improve performance in the single-container case.
\end{itemize}
Furthermore, on the experimental side, we compare the performance of more diverse containers techniques, on a wider variety of low-power IoT hardware architectures.

\section{Conclusion}

In this paper we have introduced Femto-Containers, a new architecture we desiged to enable modern DevOps on fleets of heterogeneous low-power IoT hardware.
Using Femto-Containers, authorized maintainers of IoT device software can manage (via the network) mutually isolated software modules embedded on the same microcontroller-based device.
We implemented a Femto-Container hosting engine on a common low-power IoT operating system, porting the eBPF instruction set to RIOT.
We demonstrated experimentally its performance, without requiring any specific hardware-based memory isolation mechanism, on the most common 32-bit microcontroller architectures (including Arm Cortex-M, RISC-V, ESP32).
While requiring negligible Flash and RAM memory overhead (less than 10\%),
Femto-Containers improve state-of-the-art containerization, virtualization and eBPF use on IoT microcontrollers,
by increasing security, isolation and execution speed.
In effect, Femto-Containers enables hosting (tens of) applications executing concurrently, and multiple tenants, on a single low-power IoT device.

\bibliography{bibliography}
\bibliographystyle{unsrt}

\end{document}